\documentclass{emulateapj}

\usepackage[backref,breaklinks,colorlinks,citecolor=blue]{hyperref}
\usepackage[all]{hypcap}

\begin{document}

\title{An Infrared Study of the Circumstellar Material Associated with the Carbon Star R Sculptoris}

\author{M. J. Hankins$^1$, T. L. Herter$^1$, M. Maercker$^2$, R. M. Lau$^3$, G. C. Sloan$^{4,5}$}

\altaffiltext{1}{Department of Astronomy, Cornell University, 202 Space Sciences Building, Ithaca, NY 14853, USA}
\altaffiltext{2}{Department of Space, Earth and Environment, Chalmers University of Technology, Onsala Space Observatory, 43992 Onsala, Sweden}
\altaffiltext{3}{Jet Propulsion Laboratory, California Institute of Technology, 4800 Oak Grove Drive, Pasadena, CA, 91109-8099, USA}
\altaffiltext{4}{Department of Physics and Astronomy, University of North Carolina, Chapel Hill, NC 27599-3255, USA}
\altaffiltext{5}{Space Telescope Science Institute, 3700 San Martin Drive, Baltimore, MD 21218, USA}

\begin{abstract}

The asymptotic giant branch (AGB) star R Sculptoris (R Scl) is one of the most extensively studied stars on the AGB. R Scl is a carbon star with a massive circumstellar shell ($M_{shell}\sim 7.3\times10^{-3}~M_{\odot}$) which is thought to have been produced during a thermal pulse event $\sim2200$ years ago. To study the thermal dust emission associated with its circumstellar material, observations were taken with the Faint Object InfraRed CAMera for the SOFIA Telescope (FORCAST) at 19.7, 25.2, 31.5, 34.8, and 37.1 $\mu$m. Maps of the infrared emission at these wavelengths were used to study the morphology and temperature structure of the spatially extended dust emission. Using the radiative transfer code DUSTY and fitting the spatial profile of the emission, we find that a geometrically thin dust shell cannot reproduce the observed spatially resolved emission. Instead, a second dust component in addition to the shell is needed to reproduce the observed emission. This component, which lies interior to the dust shell, traces the circumstellar envelope of R Scl. It is best fit by a density profile with $n \propto r^{\alpha}$ where $\alpha=0.75^{+0.45}_{-0.25}$ and dust mass of $M_d=9.0^{+2.3}_{-4.1}\times10^{-6}~M_{\odot}$. The strong departure from an $r^{-2}$ law indicates that the mass-loss rate of R Scl has not been constant. This result is consistent with a slow decline in the post-pulse mass-loss which has been inferred from observations of the molecular gas.

\end{abstract}


\section{Introduction}

Stars on the asymptotic giant branch (AGB) are an important source of metal enrichment and dust production in the interstellar medium (ISM) \citep[e.g.,][]{Matsuura2009,Boyer2012}. Characterizing the mass-loss of AGB stars is fundamental to our understanding of this phase of stellar evolution \citep[][and ref. therein]{Willson2000} and the evolution of the ISM in galaxies \citep{Valiante2009}. AGB stars experience helium burning in thermal pulses (TPs),  leading to the dredge-up of freshly produced carbon to their surfaces, and high mass-loss rates which transport material to the ISM \citep[][and ref. therein]{Habing1996}. Observations of circumstellar shells associated with several nearby carbon stars have provided interesting insights on mass loss during and after TP events \citep{Olofsson1990, Olofsson1996, Maercker2012}.

R Sculptoris (R Scl) is a nearby \citep[d$\sim$370 pc;][]{Maercker2016} AGB star which is well studied in a variety of different wavelengths \citep[e.g.,][]{Schoier2005,Sacuto2011}. R Scl is a carbon star with massive circumstellar shell ($M_{shell}=4.5\times10^{-3}~M_{\odot}$) that is thought to have been produced during a TP of the star $\sim$1800 years ago \citep{Maercker2012}\footnote{Note that \cite{Maercker2012} adopts a closer distance to R Scl (d$\sim290$ pc) which influences the mass and timescale determination.}. Circumstellar material associated with R Scl has been observed using scattered-light-imaging \citep{GonzalezDelgado2001,Olofsson2010}, polarization \citep{Maercker2014}, and emission from molecular gas \citep{Maercker2012,Vlemmings2013,Maercker2016}. However, the thermal dust emission is not as well studied. Previous works have analyzed mid-infrared spectra of R Scl \citep[e.g.,][]{Hony2004}, although the large radial extent of the dust shell ($R_{shell}\sim$20") presents issues with observing the total flux from the circumstellar material. 

To improve our understanding of the dust emission in the mid-infrared, R Scl was observed using the Faint Object Infrared CAmera For the SOFIA Telescope \citep[FORCAST;][]{Herter2012} at 19.7, 25.2, 31.5, 34.8, and 37.1 $\mu$m. Examining R Scl at these wavelengths is particularly interesting because it is near the peak of the observed dust emission \citep{Schoier2005}. Additionally, R Scl is known to be a carrier of the 30 $\mu$m MgS feature \cite{Hony2004} which is also of interest. The spatial resolution provided by SOFIA/FORCAST at 19.7--37.1 $\mu$m ($\sim$3.2--3.8") is sufficient to resolve the extended dust shell associated with R Scl. By modeling the spatial distribution of the dust emission, it is possible to fit the emission from the dust shell and constrain contributions from the present-day mass loss. Constraining these components provides a probe of the recent mass-loss history of R Scl. 

ALMA observations of R Scl have implied a substantial reservoir of molecular gas inside its extended shell \citep[][]{Maercker2012,Maercker2016}. The post-TP mass loss, which is responsible for generating this material, could play a substantial role in the total mass loss of R Scl, which has significant implications for its lifetime on the AGB \citep{Maercker2016}. If there is a substantial amount of dust associated with the molecular gas inside of the shell, the FORCAST observations may constrain this dust component. In either case, the FORCAST data provides an interesting comparison with the earlier ALMA observations.

In addition to studying the mass-loss history of R Scl, the FORCAST imaging observations also allow us to study the temperature and morphology of the dust emission. Previous observations of the shell have noted features which show deviations from spherical symmetry \citep{Olofsson2010,Maercker2014,Maercker2016}. Some of these features, such as the flattening in the southern portion of the shell, may be related to interactions of the circumstellar material produced by R Scl and its binary companion \citep{Maercker2014}.

\section{Observations and Data Reduction}

\subsection{SOFIA/FORCAST}

We observed R Scl with the 2.5 m telescope aboard the Stratospheric Observatory for Infrared Astronomy (SOFIA) using the FORCAST instrument \citep{Herter2012}. FORCAST is a $256 \times 256$ pixel dual-channel, wide-field mid-infrared camera with a field of view of $3.4'\,\times\,3.2'$ and a plate scale of $0.768''$ per pixel. The two channels consist of a short-wavelength camera (SWC) operating at 5 -- 25 $\mu\mathrm{m}$ and a long-wavelength camera (LWC) operating at 28 -- 40 $\mu\mathrm{m}$. An internal dichroic beam-splitter enables simultaneous observation from both short-wavelength and long-wavelength cameras, while a series of bandpass filters is used to select specific wavelengths.

FORCAST observations of R Scl were taken during SOFIA cycle 4 on 2016 July 18 and 19 on flights 323 and 324. R Scl was observed with the 19.7, 25.2, 31.5, 34.8 and 37.1 $\mu$m filters. To increase the efficiency of observations, the 19.7 \& 31.5 and 25.2 \& 34.8 were simultaneously observed using the dichroic beamsplitter, while the 37.1 filter was observed individually.

Parallel chopping and nodding on the array were used to remove the sky and telescope thermal backgrounds. The frequency of the chop throw was $\sim 4$ Hz, and the integration time at each nod position was $\sim$170 s. A five-point dither pattern aided the removal of bad pixels and mitigated response variations. Table \ref{tab:tab1} provides further details of the observations.

The final processed data of R Scl were downloaded from the SOFIA Science Center. The quality of the images is consistent with near-diffraction-limited imaging from 19.7 to 37.1 $\mu$m; the full width at half maximum (FWHM) of the point spread function (PSF) was $3.0''$ at $19.7~\mu$m, $3.1''$ at $25.2~\mu$m, $3.5''$ at $31.5~\mu$m, $4.0''$ at $34.8~\mu$m, and $4.0''$ at 37.1 $\mu$m. The RMS noise per pixel was $2.6$ mJy at $19.7~\mu$m, $3.4$ mJy at $25.2~\mu$m, $2.7$ mJy at $31.5~\mu$m, $3.9$ mJy at $34.8~\mu$m, and $2.2$ mJy at 37.1 $\mu$m. The estimated $3\sigma$ uncertainty in the flux calibration is $\pm20\%$. 

\begin{figure}[ht]
\centering
\includegraphics[width=80mm,scale=1.0]{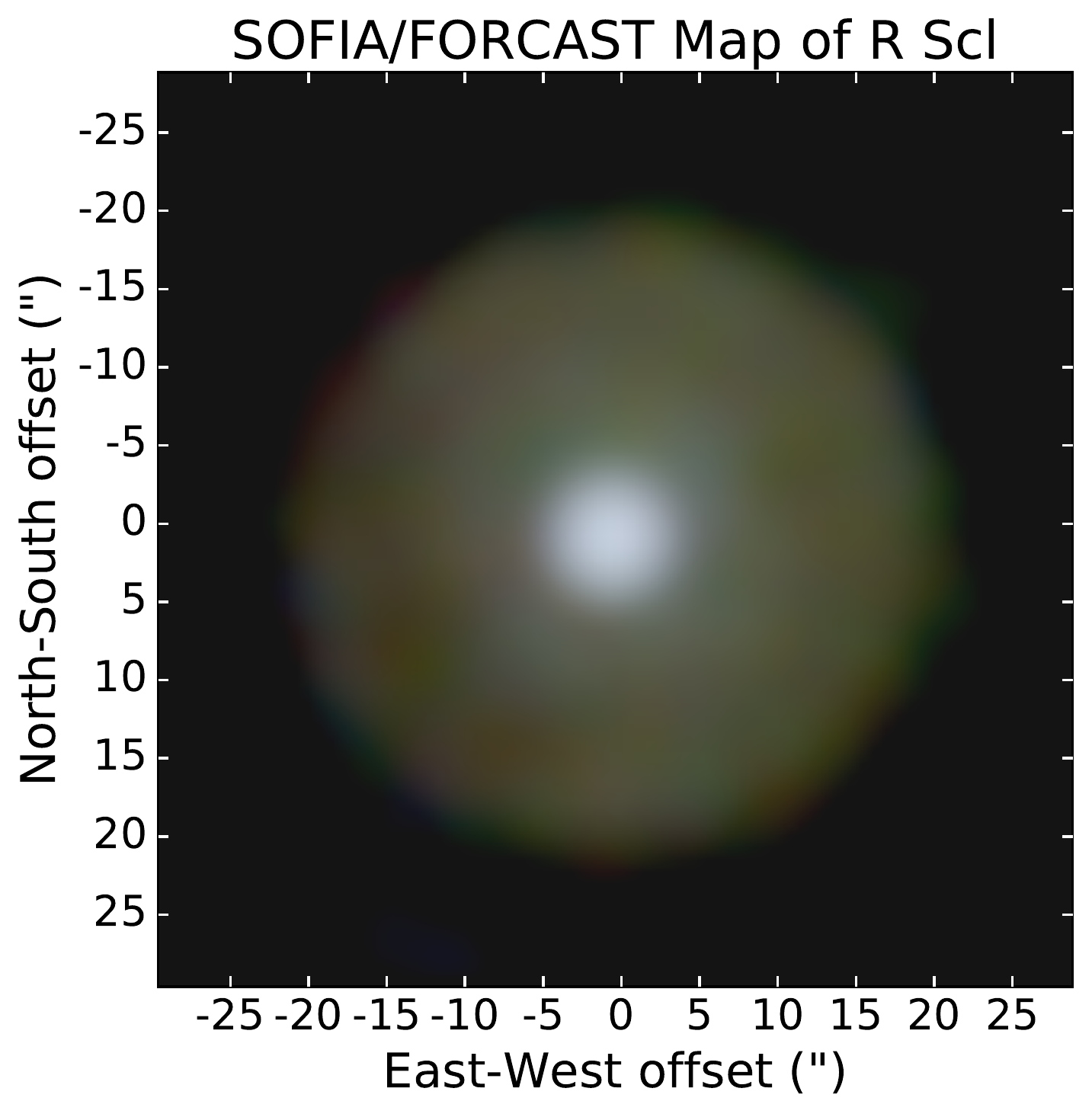}
\caption{False color map of R Scl created with the the 25.2 (blue), 31.5 (green), and 37.1 $\mu$m (red) FORCAST data.}
\label{fig:fig1}
\end{figure}

\begin{figure*}[ht!]
\centering
\includegraphics[width=160mm,scale=1.0]{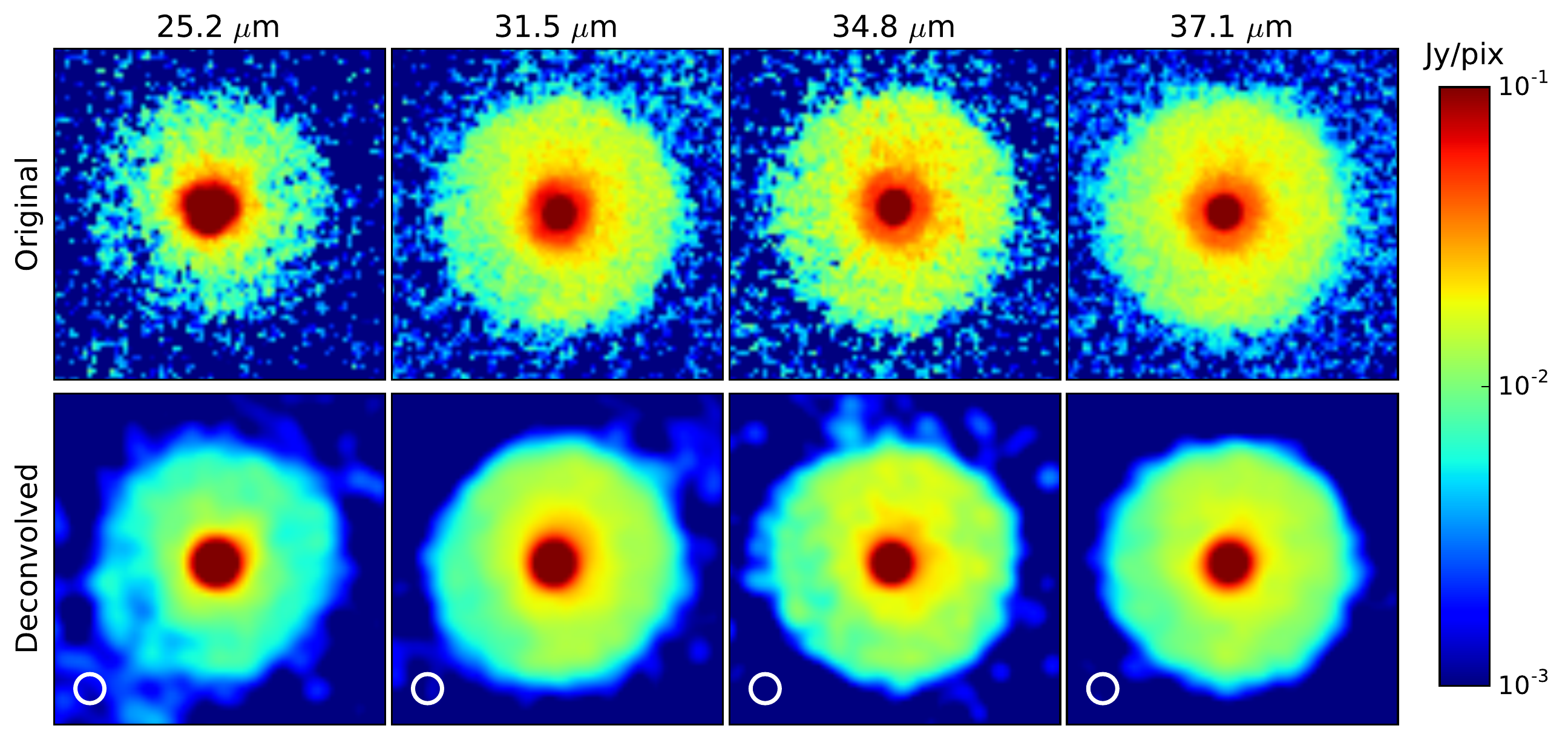}
\caption{FORCAST observations of R Scl at 25.2, 31.5, 34.8, and 37.1 $\mu$m. The top row shows the observed data, and the bottom row shows the maps after being deconvolved and convolved back to a uniform beam size. The final beam size is indicated by the white circles in the bottom left of the images. The 19 $\mu$m data is not shown in this figure because the extended emission is low signal which presents issues with the convolution.}
\label{fig:fig1N}
\end{figure*}

To bring the observations at different wavelengths to the same spatial resolution, the data were deconvolved using the Richardson-Lucy algorithm (Richardson 1972, Lucy 1974). The PSF was estimated from in flight observations of the calibrator 2 Pallas, which were taken on the same flights as R Scl. After the deconvolution, the maps were convolved back to a uniform beam size of $4.0''$. Figure~\ref{fig:fig1} displays a false-color map of R Scl using the 25.2, 31.5, and 37.1 $\mu$m data, and Figure~\ref{fig:fig1N} shows colormaps for the data at each wavelength.

\subsection{Data from Other Sources}

Data from a variety of sources were used to construct a multiwavelength SED of R Scl. Visible and near-infrared magnitudes were taken from \cite{Sacuto2011} and \cite{Whitelock2006}. Additional observations taken with Herschel/PACS at 70 $\mu$m were downloaded from the Herschel archive. These data were originally presented by \cite{Cox2012} along with Herschel observations of a large sample of AGB stars. 

Aside from photometric data, ISO/SWS spectra of R Scl were also available. These data were taken from the uniform database of ISO/SWS spectra \citep{Sloan2003}. R Scl was observed on six different occasions by ISO/SWS (Observation IDs: 24701012, 37801213, 37801443, 39901911, 41401514, 56900115). For this analysis, an average spectra was created from the different data sets. The ISO/SWS spectra are useful to compare with the FORCAST data because the wavelength coverage of the SWS usually overlap the FORCAST data. This provides us with an independent check of the FORCAST flux calibration. However, comparison between FORCAST and ISO/SWS must be treated carefully because of the extended nature of R Scl.

Aside from calibration, the spectra are also useful because they have much higher spectral resolution than the set of FORCAST filters and their wavelength coverage extends to 45 $\mu$m. Examination of the different ISO/SWS observations reveals significant variations in the Band 4 portion of the spectra (29--45.2 $\mu$m). This region of the average spectra must be treated carefully, as discussed in section 3.4.2.

\subsection{Dust Extinction}

Due to the dusty nature of R Scl, there is modest reddening of the source at visible wavelengths ($A_V=0.18$). We applied the extinction corrections from \cite{Sacuto2011} for visible and near-infrared photometry. The dust extinction at longer wavelengths is negligible. 

\section{Results and Analysis}

\subsection{Morphology of the Extended Dust Emission}

Observations of R Scl from 25.2 to 37.1 $\mu$m reveal extended dust emission around the central source.\footnote{The dust emission is also present at 19.7 $\mu$m but is only weakly detected ($\sim1\sigma$/pix).} The spatially extended dust emission at these wavelengths is somewhat uniform in surface brightness, which is a stark departure from a limb-brightened emission profile that one would expect from the dust emission of a geometrically thin shell. There is strong evidence for the presence of a dusty shell surrounding the central source at radius $R_{shell}\sim20"$ \citep{GonzalezDelgado2001,Olofsson2010,Maercker2014}. The temperature of the shell \citep[T$_{shell}=75$ K; ][]{Schoier2005} indicates that it should contribute significantly in the mid-infrared. The fact that the observed mid-infrared emission does not appear limb-brightened like a shell is intriguing.

To study the extended infrared emission in greater detail, radial profiles of the emission were created using the 25.2, 31.5, 34.8 and 37.1 $\mu$m maps. The average radial profiles were generated using an azimuthally averaged radial profile over the map, and statistical uncertainties were determined from the sample of pixels at a given radius. Using the average 37.1 $\mu$m profile, the radial size of the dust emission was measured at 19.2". Compensating for the convolved beam size (FWHM$\sim$4.0"), this implies a deconvolved radius of 18.8". This measurement is consistent with the radius of the dust shell determined by \cite{Olofsson2010} where $R_{shell}=$18.5--19.9".

\begin{figure}[ht]
\centering
\includegraphics[width=80mm,scale=1.0]{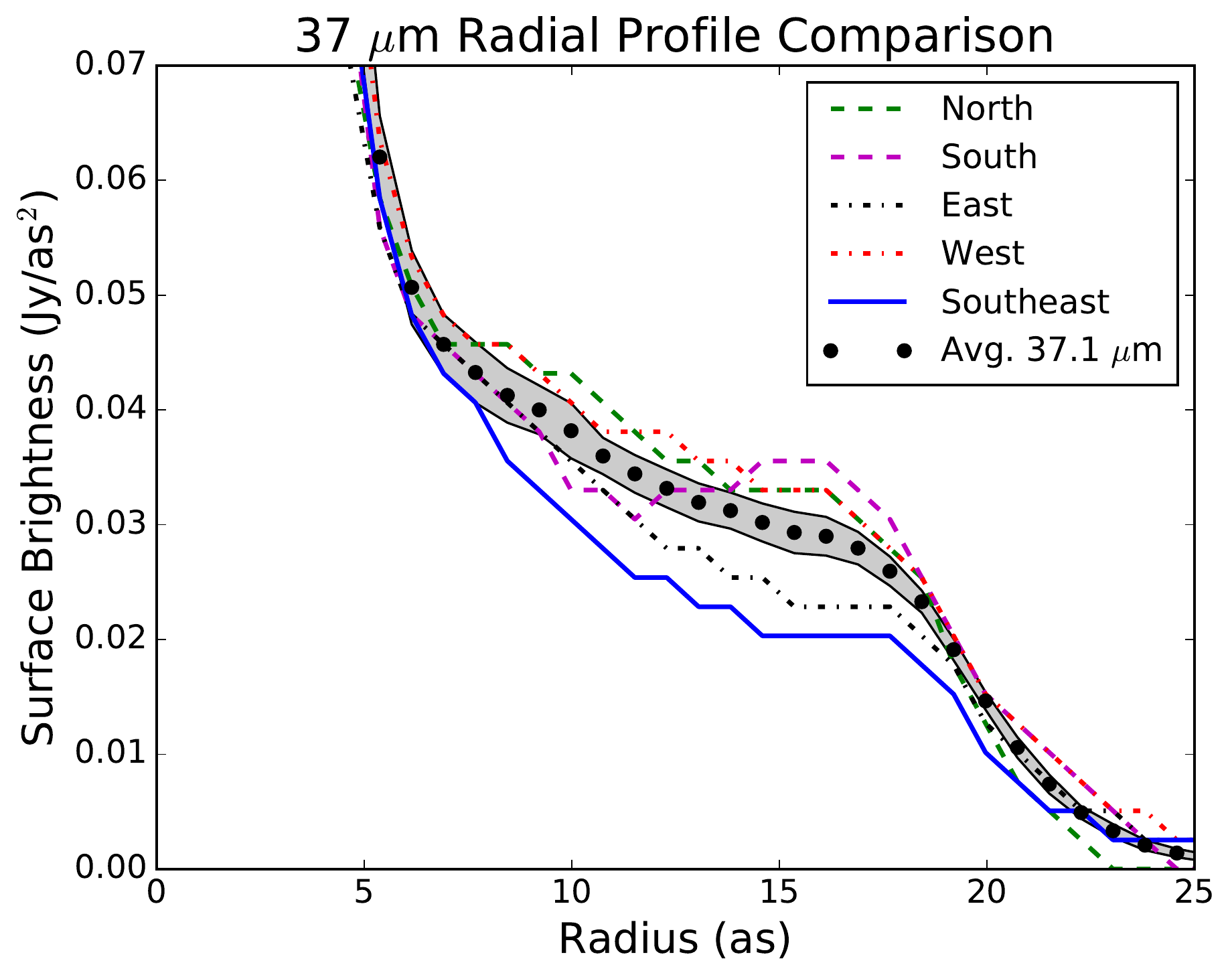}
\caption{A comparison of the average radial profile at 37 $\mu$m with cuts taken through different portions of the data, as indicated in the legend. An additional cut is shown in the southeast, which shows the strongest drop in the observed emission.}
\label{fig:fig2N}
\end{figure}

Further examination of the extended dust emission reveals an interesting dimming toward the east of the central source, which is apparent in each of the FORCAST maps from 25.2 to 37.1 $\mu$m. To show this feature more clearly, the average radial profile at 37.1$\mu$m was plotted along with a several line cuts through the map in Figure \ref{fig:fig2N}. The dip in the emission is strongest toward the southeast. It is interesting to note that this feature in the east/southeast corresponds to a region that appears dim in scattered light, but bright in polarization \citep{Maercker2014}. These characteristics could indicate a temperature or optical depth effect, though there are several other possible explanations. The next section examines the temperature structure of the dust emission.

\subsection{Color-Temperature Map}

\begin{figure}[ht]
\centering
\includegraphics[width=80mm,scale=1.0]{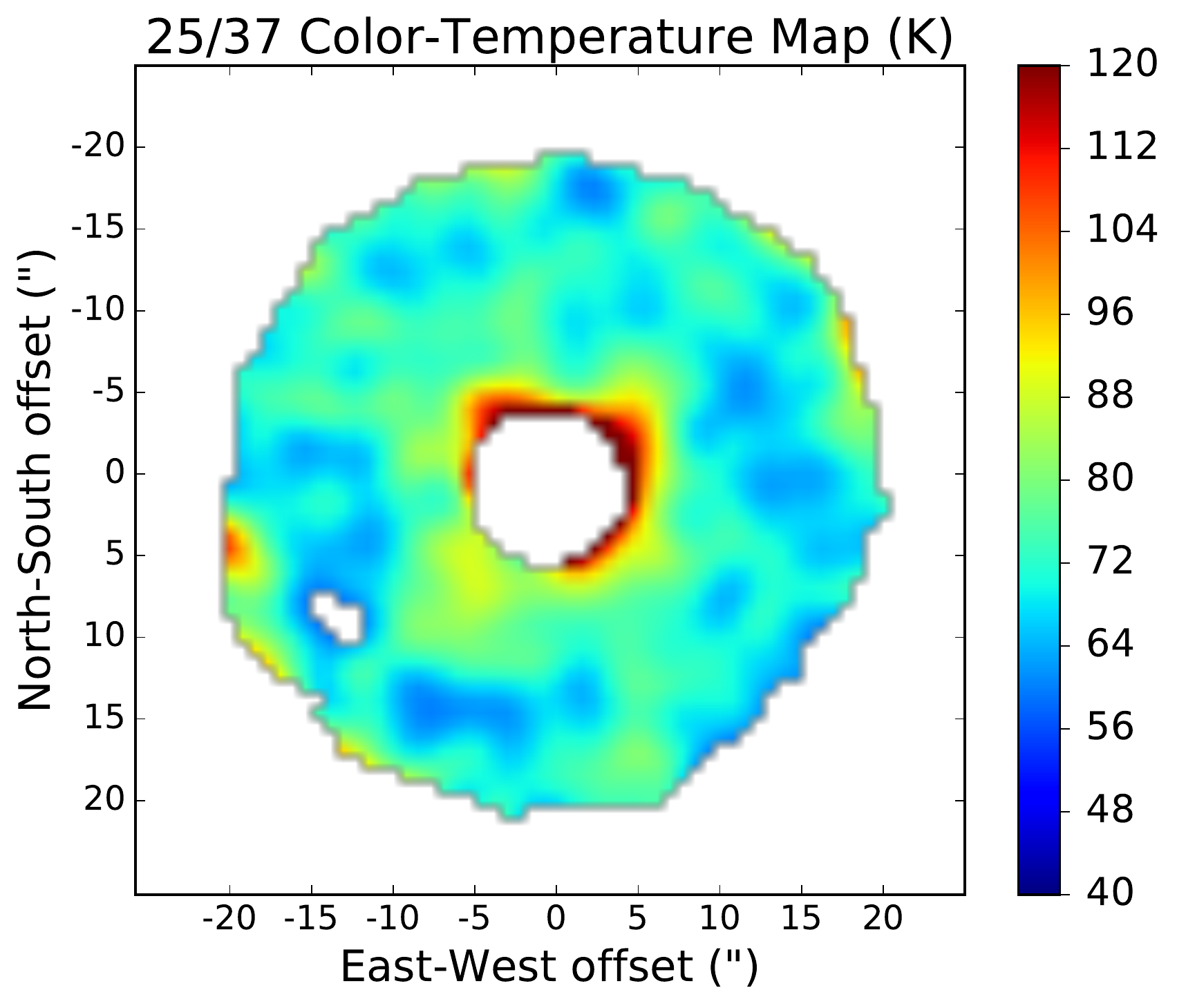}
\caption{Color-temperature map of the thermal dust emission associated with R Scl, generated using 25 and 37 $\mu$m FORCAST data. The bright central source has been masked to better show the extended dust features. Additional details can be found in \S3.2. The average temperature of the dust emission is 74$\pm$10 K.}
\label{fig:fig2}
\end{figure}

To study the temperature of the dust associated with R Scl, we created a color-temperature map of the observed emission. We assumed that the dust emission is optically thin and can be expressed as $F_{\nu} \sim Q_{\nu} B_{\nu}(T_d)$, where $Q_{\nu}$ is the optical efficiency of the dust and $T_d$ is defined as the color-temperature. Since $Q_{\nu}\propto \nu^{\beta}$ (or equivalently $Q_{\nu}\propto \lambda^{-\beta}$), the proportionality of the flux with frequency is simply: $F_{\nu} \propto B_{\nu}(T_d) \nu^{\beta}$. Taking a ratio of two filters leaves $T_d$ and $\beta$ as the only unknowns. Adopting an index ($\beta=2$), the color-temperature can be calculated (Figure \ref{fig:fig2}). 

The average temperature of the dust emission is $T_{d}$=74$\pm$10 K.\footnote{There is an additional 15\% uncertainty in the absolute value of the color-temperature due to the absolute flux calibration of the two bands.} This result corresponds well with a grey-body peaking near $\sim30~\mu$m and agrees with a previous determination of the temperature of the dust from fitting the SED \citep[$T_d=75\pm15$ K;][]{Schoier2005}. Overall the temperature of the emission appears fairly uniform, though this may be slightly deceptive since the emission has been collapsed along the line of sight. If there are multiple dust components along the line of sight, it would not be possible to separate them in a map like this. Additionally, the broad MgS feature located near 30 $\mu$m could influence the temperature determination. While the 25 $\mu$m filter is likely unaffected by this feature, the FORCAST bands at 31, 34, and 37 $\mu$m may have some additional contribution from MgS dust. Section 3.4 will return to this point with more sophisticated models of the dust emission.

\subsection{Infrared Spectra of R Scl}

\begin{figure}[ht]
\centering
\includegraphics[width=80mm,scale=1.0]{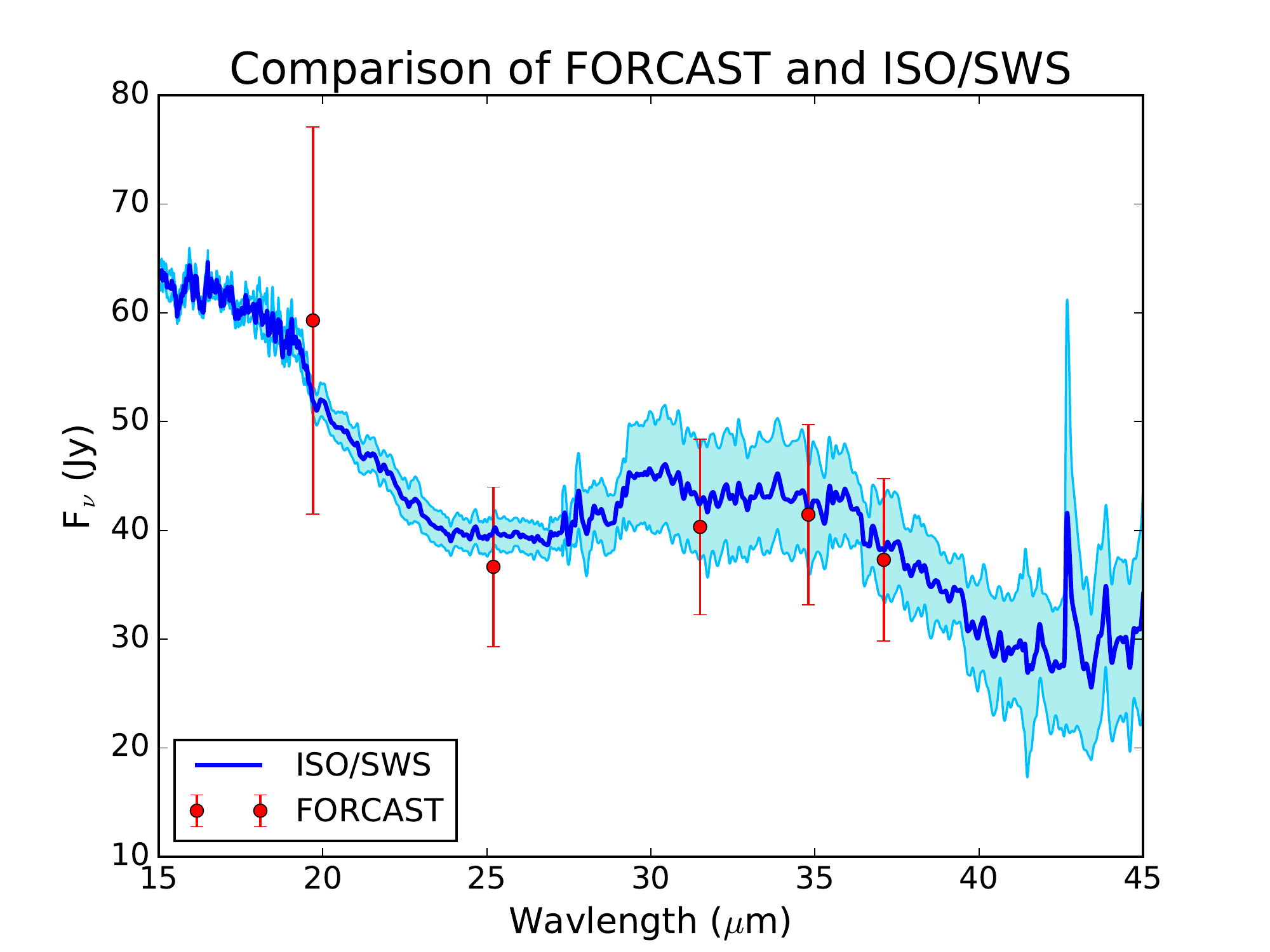}
\caption{Comparison of the average ISO/SWS spectra of R Scl with the observed FORCAST photometry, which has been extracted in an aperture matching the ISO/SWS observations. The shaded regions of the spectra indicate the uncertainty due to source variability \citep[e.g.,][]{Onaka2002} and mismatch problems between the different ISO/SWS bands \citep[see][]{Sloan2003}.}
\label{fig:fig4}
\end{figure}

ISO/SWS spectra of R Scl provide a large range of contiguous wavelength coverage (2.4-45$\mu$m), which overlaps with FORCAST, and moderate spectral resolution \citep[R$\sim$930-2450;][]{Leech2003}. To account for the extended nature of R Scl ($R_{shell}\sim20"$), which is much larger than the ISO/SWS aperture sizes (14"$\times$20" to 20"$\times$30"), photometry from the FORCAST maps were extracted over apertures equivalent to the ISO/SWS. Figure \ref{fig:fig4} plots these data, along with the average spectra of R Scl. Overall, FORCAST photometry and the ISO/SWS spectra agree fairly well.  This demonstrates consistency between the two data sets over the range from 19.7 to 37.1 $\mu$m.

By comparing the fraction of the flux contained within the ISO/SWS aperture to the total source flux observed by FORCAST at a given wavelength, it is possible to derive correction factors for the ISO spectra as if it had observed the full source. For the different FORCAST filters, the fractional flux contained in the ISO/SWS aperture is 1.0 at 19.7 $\mu$m, 0.65 at 25.2 $\mu$m, 0.59 at 31.5 $\mu$m, 0.58 at 34.8 $\mu$m, and 0.59 at 37.1 $\mu$m. For comparison, the ratio of the area of the largest ISO/SWS aperture (20"$\times$33", corresponding to 660 arcsec$^2$) and the extended dust emission (R=18.8", corresponding to 1110 arcsec$^2$) is 0.6. Assuming that the flux from the source is uniformly distributed, this value agrees well with the ratios derived from the FORCAST observations, except at 19.7 $\mu$m where the flux from the central point source is more dominant.

\subsection{DUSTY Models}

The radiative transfer code DUSTY \citep{Nenkova2000} was used to model the thermal dust emission. For an initial model, we started with a simple geometrically thin dust shell. Parameters for the shell, such as the mass, size, and temperature, were adopted from the literature. Table \ref{tab:tab2} lists these values and their references. To improve the model, these parameters were allowed to vary and additional dust components were also considered. This process is described in the following subsections. For each model, DUSTY was used to compute the expected SED for the source as well as radial profiles for each of the FORCAST bands. These results were compared to the observed data using a $\chi^2$ analysis to determine the best-fitting model.

\subsubsection{Dust Properties}

In the DUSTY models, an MRN size distribution was adopted for the dust grains \citep[$dn/da\propto a^{-3.5}$, $a_{min}=0.005~\mu m$, and $a_{max}=0.25~\mu m$;][]{Mathis1977}. While it is possible to modify the grain sizes, for ISM grains with typcial sizes, the thermal emission is relatively insensitive to this effect. The dust composition was modeled as a mixture of three dust species: amorphous carbon (AmC), silicon carbide (SiC), and magnesium sulfide (MgS). The fractional abundance by mass for each component was 0.86 AmC, 0.10 SiC, and 0.04 MgS. These values are consistent with the abundances of SiC and MgS determined by \cite{Sacuto2011} and \cite{Hony2004}. Optical properties for each dust component were taken from \cite{Zubko2004}, \cite{Pegourie1988}, and \cite{Begemann1994}. 

AmC dust dominates the total dust mass, and uncertainties in the optical constants for this dust species can have a substantial impact on various model parameters. For example, Groenewegen \& Sloan (2017) find that models of carbon stars using the optical properties of \cite{Zubko1996} result in dust masses which are a factors of $\sim$5-11 times less than models using the optical properties of \cite{Rouleau1991}. For this work, the optical constants from \cite{Zubko2004} were adopted because they are able to reproduce physical properties of the dust shell such as the size and temperature. Models with the constants from \cite{Rouleau1991} cannot reproduce the observed size and temperature of the shell without forcing the luminosity to unrealistically low values. In stating this, it is important to keep in mind that model-dependent uncertainties in the total dust mass have a substantial effect on calculating the total mass loss.

Studies of MgS dust in AGB stars have found that a cumulative distribution of ellipsoidal particles (CDE) does a better job of reproducing the observed 30 $\mu$m feature than a distribution spherical particles \citep{Begemann1994,Hony2004}. Consequently, the optical properties from \cite{Begemann1994} were used to calculate extinction coefficients for CDE shapes according to \cite[][chp. 12]{Bohren1983}. To examine the effect of the MgS dust on the spectrum of R Scl, models with varied MgS abundance from 0 to 10\% were studied. The upper limit on the dust abundance is set by the limited amount of Mg and S available to produce dust \citep{Hony2004}.

\subsubsection{Geometry}

Figure \ref{fig:fig5} panel A shows the radial emission profile for the initial shell model along with the observed radial profile at 37 $\mu$m. The data show a clear excess at radii between ($\sim5-15"$) compared to the model. While it might be possible to make slight improvements to the fit of this model by scaling up the dust mass, the shape of the profile, which is largely dictated by the geometry of the shell, cannot reproduce the observed radial profile. Thus, the observed emission cannot be explained by the shell model alone. As noted in \S3.1, the dust shell is expected to be dominant in mid-infrared wavelengths. Therefore, the excess dust emission is likely a secondary dust component, which appears to be located interior to the shell.

To study the excess emission, the model from panel A was subtracted from the observed radial emission profile. Panel B of Figure \ref{fig:fig5} shows the resulting profile. Since the dust surrounding R Scl is sufficiently optically thin, it is possible to model the secondary dust component separate from the shell component. To model the excess emission, DUSTY was used to create an additional set of models with different radial density distributions for the dust. For the models, a power law was adopted for the density profile with $n\propto r^{\alpha}$, and varied the value of $\alpha$ from -2 to 2 in increments of 0.25. Additional model parameters, such as the inner and outer radii, were also varied to ensure goodness of fit. Table \ref{tab:tab2} lists these values and additional parameters used in the models. 

\subsubsection{Results}

\begin{figure}[ht]
\centering
\includegraphics[width=80mm,scale=1.0]{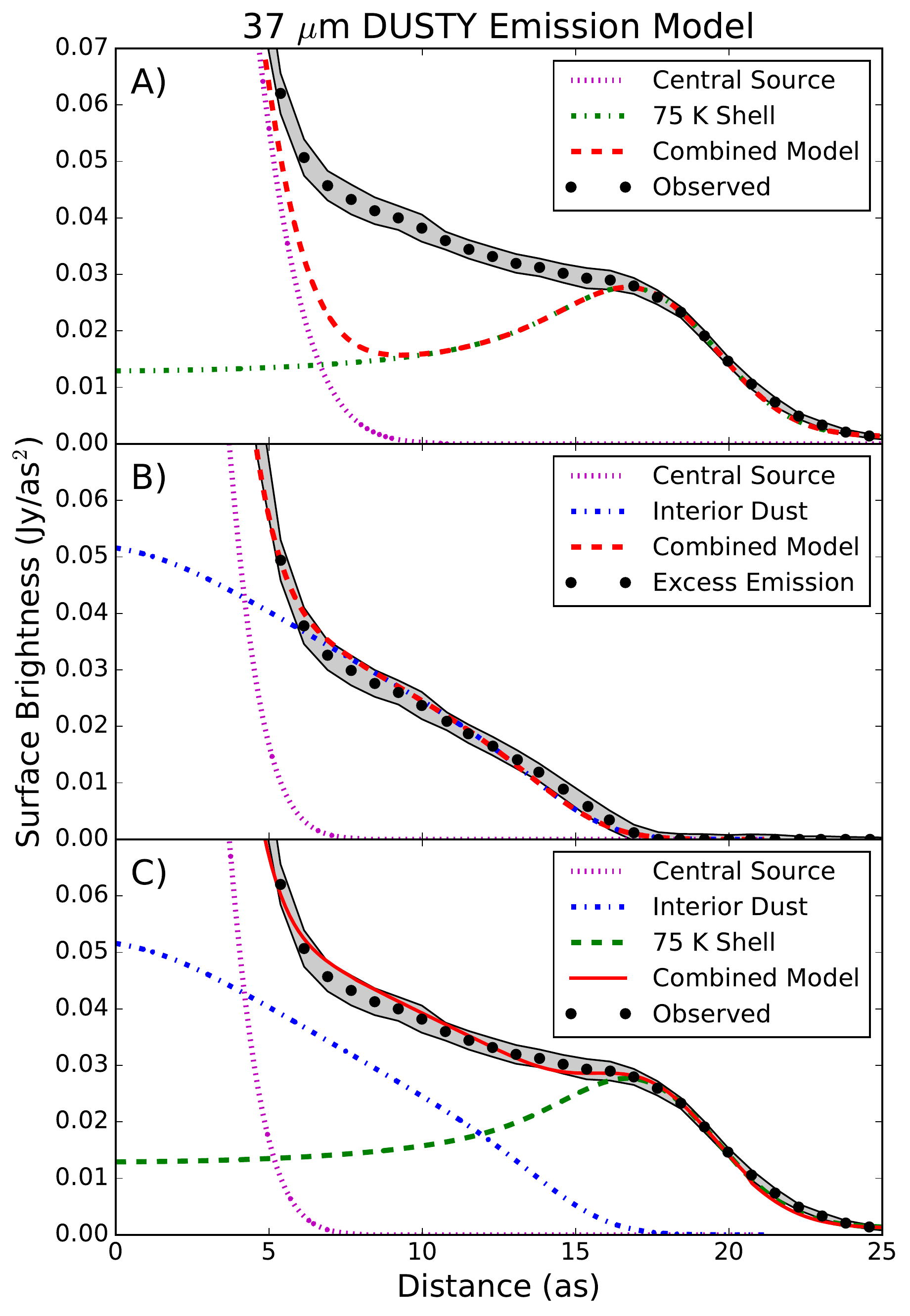}
\caption{Observations of the radial emission profile of R Scl at 37.1 $\mu$m along with models of the dust emission generated by DUSTY. Panel A shows the observed radial profile along with a model for the dust emission from the shell surrounding R Scl. The model for the shell is a poor fit to the data. Panel B shows the residual emission left after subtracting the shell model from panel A. This residual was fitted to determine the radial density profile of the interior dust component. Panel C shows the combination of the best-fit model from panel B along with the shell model from panel A. Note that the observed profiles only consider statistical uncertainties in the data and not the absolute flux calibration.}
\label{fig:fig5}
\end{figure}

\begin{figure}[ht]
\centering
\includegraphics[width=80mm,scale=1.0]{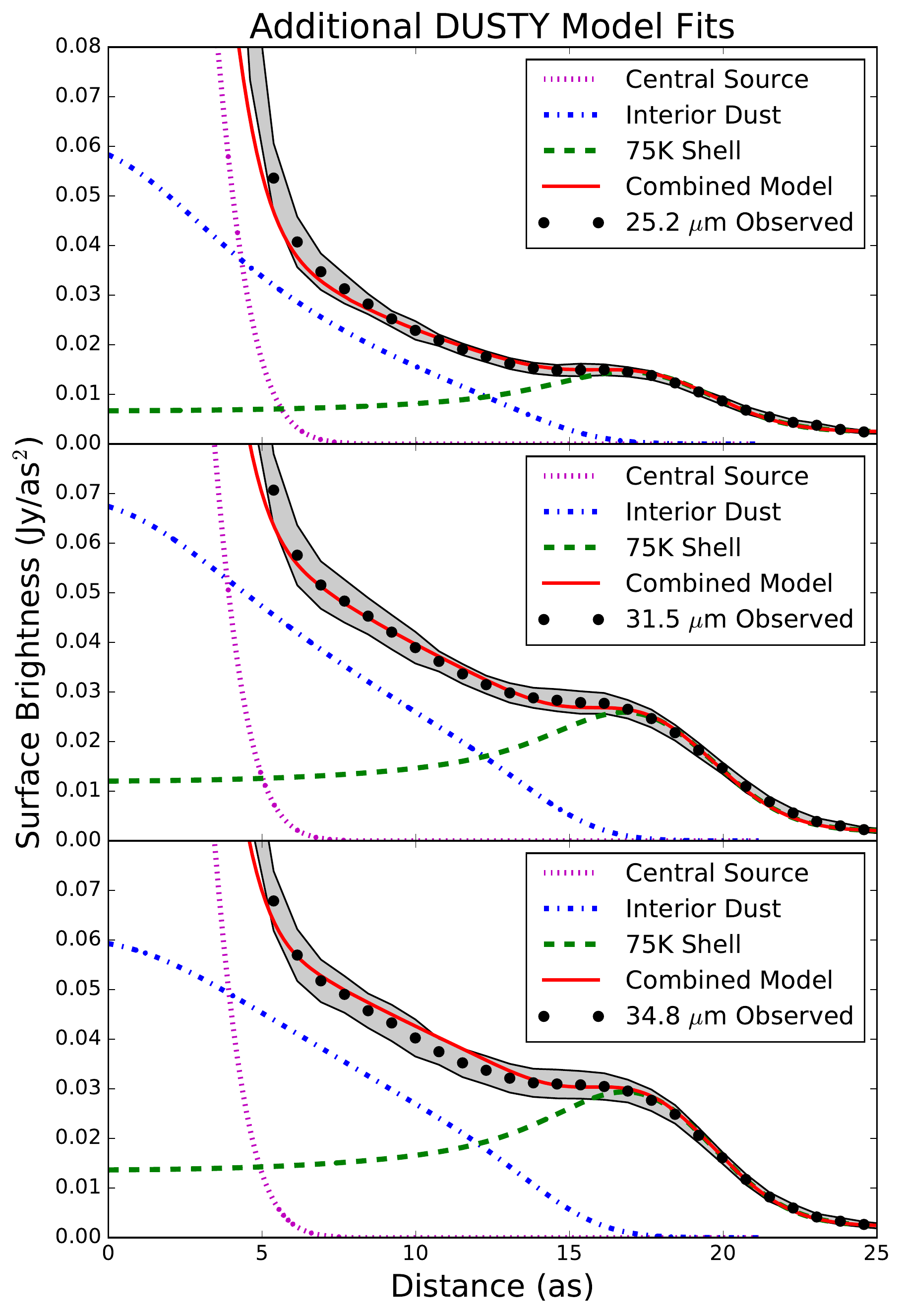}
\caption{Observations of the radial emission profile of R Scl at 25.2, 31.5, and 34.8 $\mu$m. Models for the shell and interior dust component are shown for each wavelength. The two components combined do a excellent job of reproducing the observed radial emission profile at each wavelength. Note that the observed profiles only consider statistical uncertainties in the data and not the absolute flux calibration.}
\label{fig:fig6}
\end{figure}

The best-fit radial density profile for the interior dust component was found to be $\alpha=0.75^{+0.45}_{-0.25}$. Panel C of Figure \ref{fig:fig5} shows the combined model with the emission from the shell and the interior dust distribution. Corresponding models for each of the other FORCAST wavelengths were compared with the observed data to ensure consistency. Figure \ref{fig:fig6} shows these models and data agree well for the other three bands.

\begin{figure}[ht]
\centering
\includegraphics[width=80mm,scale=1.0]{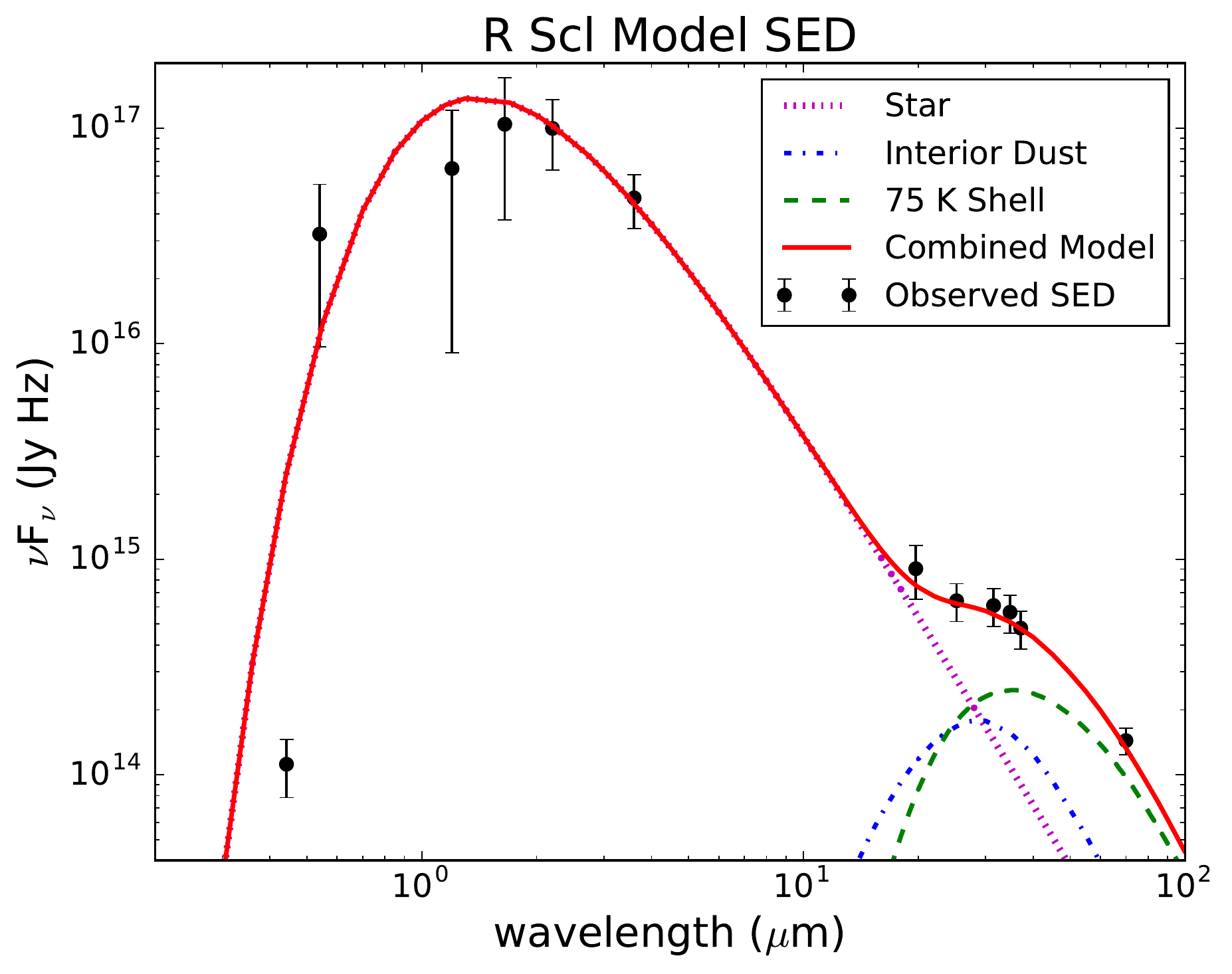}
\caption{Observed SED of R Scl with data taken from \cite{Sacuto2011,Whitelock2006,Cox2012} and the FORCAST photometry presented in this work. The best-fit two-component (shell + interior dust distribution) DUSTY model determined in section 3.4 is shown for comparison.}
\label{fig:fig7}
\end{figure}

Next, the model SED of R Scl was compared with the observed photometry to ensure the different physical parameters adopted for the stellar component and the dust at longer wavelengths agree with the data. Figure \ref{fig:fig7} plots the multi-component DUSTY model and the observed SED of R Scl. Overall, the fit for the DUSTY model does a satisfactory job of reproducing the observed SED.  

Last, the dust mass was calculated for the DUSTY model at different wavelengths to check the overall scaling to the observed flux. The dust mass for the 37.1 $\mu$m model is $M_d=9.0^{+2.3}_{-4.1}\times10^{-6}~M_{\odot}$. The model at 31.5 \& 34.8 $\mu$m are both comparable to this, while the 25.2 $\mu$m is slightly smaller ($M_d\sim4\times10^{-6}~M_{\odot}$). This difference could be related to the broad MgS feature because the 31.5, 34.8, and 37.1 $\mu$m bands likely have some contribution from it. Alternatively, the difference in the masses could indicate deviations from the expected radial temperature profile. Determining the dust mass is strongly dependant on temperature and could easily explain the observations. In either case, these effects are difficult to infer with the available data.


Considering the dust abundances further, it is challenging to place constrains on the strength of the MgS emission feature using the FORCAST photometric data alone. The level of the feature on top of the continuum emission appears comparable to the FORCAST calibration uncertainty. To improve constraints on the models, the ISO/SWS spectra were also examined. The broad 30 $\mu$m feature is more clearly visible in the average spectra of R Scl because of the increased wavelength coverage and higher spectral resolution. Although, the shape and strength of the feature between the six different ISO observations shows significant variation. These changes are likely artifacts, as there are well documented issues with hysteresis in band 4 of the SWS \citep[ranging from 29--45 $\mu$m; ][]{Sloan2003}. Thus, the data in this region must be treated with considerable caution.

Section 3.3 has already compared the ISO/SWS spectra and FORCAST photometry to show the agreement between the data from 19.7 to 37.1 $\mu$m. However, the spectrum beyond this is suspicious. Comparing SED models of R Scl to the spectra, the downturn in the spectra from $\sim$ 35 -- 45 $\mu$m cannot be reproduced while also fitting the photometry out to 70 $\mu$m. It is unfortunate that the spectra at these wavelengths are not more reliable, as it would greatly improve constraints on the MgS dust emission. Considering these issues, it is not possible to provide updated constraints on the MgS abundance with the available data.

\section{Discussion}
 
\subsection{Mass-loss History}

By constraining the density profile of the interior dust distribution, the FORCAST data have provided us with a glimpse into the post-TP mass loss of R Scl. The following steps were taken to model the mass-loss history for R Scl. First, the radial distribution of dust was used to infer the amount of mass at a given radius from R Scl. This depends on both the density profile (fitted in \S3.4.1) and the gas-to-dust ratio of the material \citep[$\sim$590;][]{Schoier2005}. Next, the timescale of material at different radii was computed based on its expansion velocity. The observed velocity of material associated with R Scl varies from $\sim$14.5 km/s, for the shell component, to $\sim$10.5 km/s, for its present day wind \citep{Maercker2012}. This variation in gas velocity was approximated as a linear increase of velocity with radius. With the FORCAST data, which does not provide any velocity information, it is not possible to more accurately model the expansion of the circumstellar material.

\begin{figure}[ht]
\centering
\includegraphics[width=90mm,scale=1.0]{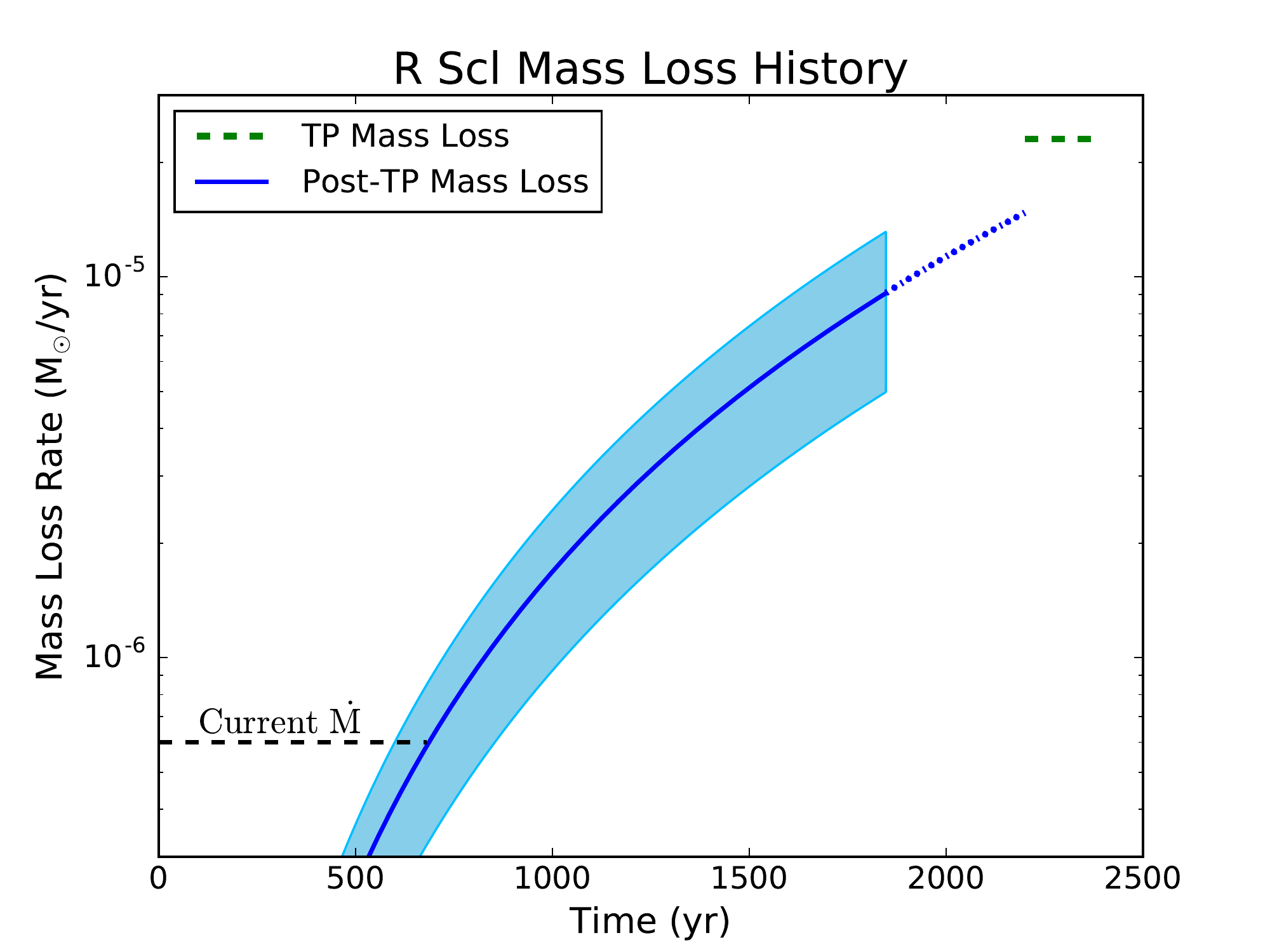}
\caption{Mass-loss history for R Scl inferred from the density profile of the interior dust component (blue line). The uncertainty in the mass loss rate (light blue region) was estimated from the uncertainty in the total dust mass determined by the DUSTY models. As discussed in the section 4.1, there are several additional uncertainties in determining the absolute mass-loss rate which make the value difficult to constrain. However, the relative change in the mass-loss rate is well constrained from modeling the extended dust emission. The blue dotted line is an extrapolation of the inferred mass loss rate out to the time when the TP occurred, and the black dashed line indicates the current mass loss rate.}
\label{fig:fig9}
\end{figure}

Figure \ref{fig:fig9} shows the resulting model of the mass-loss history of R Scl. The solid blue curve denotes the expected mass loss history from the best-fit model, and the green dashed line denotes the mass loss rate of R Scl at the time of the thermal pulse \citep{Maercker2016}. Because the FORCAST observations are limited by their spatial resolution ($\sim$4''), some regions in the plot should be taken with some degree of skepticism. For example, recently produced material ($t\lesssim400$ yr) is not well constrained by the FORCAST data because its emission is blended with the bright central source. Additionally, the transition region between the interior dust distribution and the outer shell is poorly constrained. The shell component appears to dominate the emission from R Scl at its outermost radii. However, some interior dust may be blended with and contribute to the emission here as well. High-resolution observations of the circumstellar dust at visible and near-infrared wavelengths may help to constrain the dust components in this region.

Using the resulting mass-loss history from the interior dust component, the average post-TP mass-loss rate was found: $\dot{M}\sim2.3\times10^{-6}~M_{\odot}$/yr. This value is uncertain due to uncertainties in the gas-to-dust ratio, the expansion velocity of the material, and the total dust mass. Considering these issues, it is important to note that while the absolute mass-loss rate may have large uncertainties, the relative change in the mass-loss rate is well constrained by the fitted radial density profile.

Observations of molecular gas emission from R Scl have also been used to study material inside of the shell.\footnote{This material inside of the shell is referred to as the circumstellar envelope (CSE) in \cite{Maercker2012,Maercker2016}, which is equivalent to the interior component studied in this work.} Using these data, \cite{Maercker2016} infer an average post-TP mass-loss rate of $\sim1.6\times10^{-5}~M_{\odot}$/yr. Because of the different methods for determining the mass-loss rate and the large uncertainties associated with each method, it is difficult to compare the mass-loss rates from this work and \cite{Maercker2016}. 

The expected post-TP mass loss for R Scl based on the classical TP scenario \citep[$\sim7\times10^{-3}~M_{\odot}$; ][]{Maercker2016} is comparable to the mass of the interior component determined by the FORCAST observations ($\sim5\times10^{-3}~M_{\odot}$). However, the classical scenario for TPs assumes a sharp decline in the mass loss after the TP, which is inconsistent with the model presented here. The slow declining behavior in the evolution of the mass loss rate is more similar to the result found by \cite{Maercker2012,Maercker2016}.

Considering the slow decline in post-TP mass-loss rate of R Scl merits some discussion on TPs and other carbon stars with high mass-loss rates. Other well-known carbon stars are known to have high and unsteady mass-loss rates \citep[e.g. IRC +10216;][]{Sloan1995,Decin2011}. While these more extreme objects may be unrelated to normal TP phenomena, more typical post-TP objects can also show multiple shells, which indicate strong variations in mass-loss rate \citep[e.g., U Ant;][]{GonzalezDelgado2003}. Studies of additional objects will help determine if R Scl represents the larger population of typical carbon stars. Clearly, more work can be done to address these issues.

\subsection{Comparisons with Polarized Light Observations}

Observations of scattered light from dust at visible and near-infrared wavelengths have been used to infer the distribution of dust associated with R Scl \citep{GonzalezDelgado2001,Olofsson2010}. Examining the scattered-light intensity is different from the thermal emission because the scattered-light signal is not dependant of the temperature of the dust. Nonetheless, these two types of observations have a complementary nature because they both trace the dust distribution. Previous models of the spatial distribution of scattered-light have mainly considered effects from the shell \citep[e.g., ][]{Maercker2014}. However, the interior dust distribution discussed in this work could impact these observations. We modeled the light scattering and polarization for R Scl to examine this effect.

For the model, the geometric parameters for the shell and the interior dust distribution were adopted from the models in \S4.3, with the exception that the interior dust distribution was allowed to extend out to a radius of $\sim$18", which terminates at the shell component. To calculate the intensity profile of scattered light, a simple single scattering model was assumed. The integrated scattered-light intensity was calculated along a path, $s$, described by an impact parameter $b$, and radius $R$, which are related: $b=\sqrt{R^2+s^2}$. To estimate the polarized light contribution, a Rayleigh polarization model was used  ($P(\theta)=(1-\cos(\theta)^2)/(1+\cos(\theta)^2)$), along with the phase function from \cite{Draine2003}\footnote{Note that for $\delta$=0, this expression reduces to the Henyey-Greenstein phase function}. Last, the computed model profiles were convolved with a Gaussian with FWHM$\sim1.3$" to match the observed polarization data.

\begin{figure}[ht]
\centering
\includegraphics[width=80mm,scale=1.0]{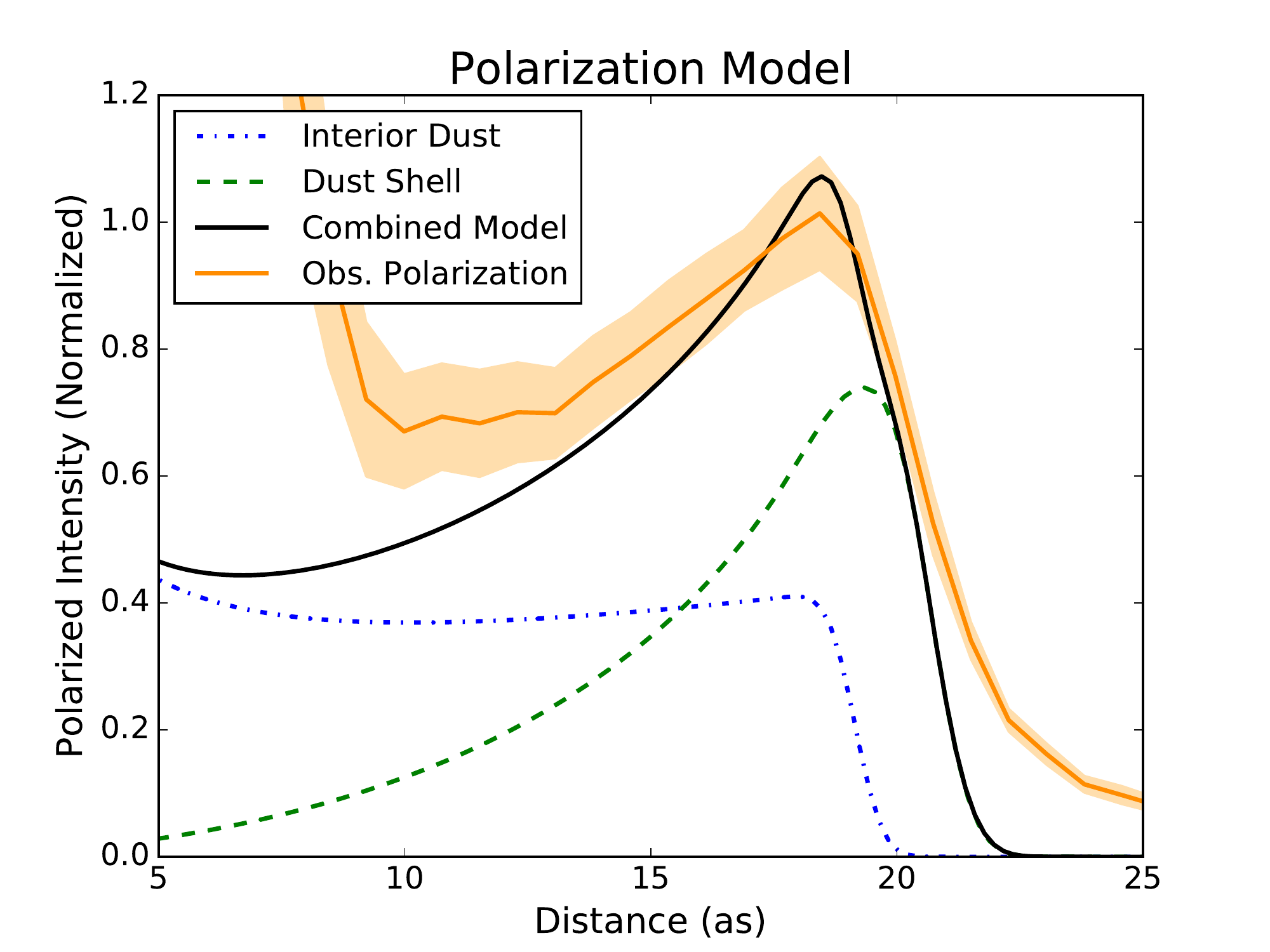}
\caption{The polarized-light model for the dust shell and interior dust distribution is shown (black) along with the observed R-band polarization from \cite{Maercker2014} (orange). Considering the effect of an interior dust distribution in addition to the shell fits the data better than the shell alone. The data interior to $\sim10$" is not reliable due to an issue with the coronagraph.}
\label{fig:fig10}
\end{figure}

Figure \ref{fig:fig10} shows the model for the polarized light intensity along with the R-band polarized light observations of R Scl from \cite{Maercker2014}. The combined model (shell + interior dust) fits the observed polarization better than the shell alone. Previous works, which only considered a single dust shell, underpredicted the amount of polarized light coming from radii interior to the shell. This may have signaled the existence of a substantial interior dust component. However, the quality of the polarization data, which had issues with the central source masked by the coronograph ($\lesssim$10"), made this emission somewhat suspect. Now that the thermal dust emission has clearly shown the presence of an interior dust component, the results of these two data sets appear consistent.  Improved polarization data with suppressed effects from the chronograph may provide even better characterization of the interior dust distribution and properties of the dust grains.

\subsection{Detailed Morphology: Evidence of Binary Interaction?}

The structure and morphology of the circumstellar material associated with R Scl may provide insights into the nature of the source. The dimming in the emission toward the east/southeast discussed in \S3.1 is a particularly interesting feature. The physical cause of this feature is unclear, though examining this region at other wavelengths may provide some insight. In the scattered light observations of R Scl, the eastern portion of the emission also appears somewhat fainter than other regions \citep{Olofsson2010}, though it appears bright in polarized light \citep{Maercker2014}. These characteristics could indicate the dust in this region is cooler and denser than other parts of the circumstellar material. 

Returning to the color-temperature map (Figure \ref{fig:fig2}), the derived dust temperatures do not show large variations. However, part of the region to the southeast falls below the signal-to-noise cutoff at 25.2 $\mu$m. In principle, this could be caused by emission which is much cooler than its neighboring regions. Referring to the 70 $\mu$m maps from Herschel, the region to the southeast still appears dimmer than its neighboring regions. The southeast feature is not likely to result from changes in temperature because the emission at long wavelengths should appear more comparable to neighboring regions, but it could be indicative of a lower dust column along this line of sight.

 The flattened southern portion of the shell may have some relation to the southeast feature. \cite{Maercker2016} suggest that it may have resulted from an interaction with R Scl's binary companion shortly after it was produced. Could an interaction between material produced by R Scl and its binary companion also be responsible for the relatively dim regions in the mid-infrared? It is well established that R Scl's binary companion is acting to shape the circumstellar material around R Scl, as evidenced by the interesting spiral-like morphology observed in CO emission \citep{Maercker2012}. However, it is unclear if these interactions could also produce a much larger-scale asymmetry in the circumstellar material. Detailed hydrodynamical modeling of the motions of the gas and dust in the circumstellar environment of R Scl are needed to improve our understanding of this system.

\section{Conclusions}

We studied the thermal dust emission from the AGB star R Scl using observations from SOFIA/FORCAST at 19.7, 25.2, 31.5, 34.8, and 37.1 $\mu$m. These mid-infrared imaging observations show spatially extended dust emission associated with the circumstellar environment of R Scl. Radiative-transfer modeling with DUSTY revealed that the observed spatial distribution of the dust emission cannot be replicated by a shell alone. Instead, the observed spatial distribution of emission can be fitted by a shell with an interior dust component. This interior component is best-fitted by a power law $n \propto r^{\alpha}$ with $\alpha=0.75^{+0.45}_{-0.25}$, which gives a dust mass of $M_d=9.0^{+2.3}_{-4.1}\times10^{-6}~M_{\odot}$. The strong departure from an $r^{-2}$ density profile indicates that the mass-loss rate of R Scl has decreased since the last TP but it has not fallen off as quickly as expected. This result is consistent with a slow decline in the post-TP mass-loss rate inferred from observations of the molecular gas.

The average post-TP mass-loss rate estimated from the dust emission ($\sim2.3\times10^{-6}~M_{\odot}$/yr) is comparable to the post-TP mass-loss rate inferred from CO observations of R Scl, although the mass-loss rate inferred from the dust emission is more consistent with the post-TP mass loss expected from the classical TP scenario. However, the fitted density profiles indicate a slow decline in the post-TP mass-loss rate which differs markedly from classical models of TPs. This result is consistent with findings from \citep{Maercker2012} which also infer a slowly declining post-TP mass-loss rate. In a broader context, this finding may have important implications on the mass loss of AGB stars during TP cycles, which greatly impacts the lifetime of stars during this phase \citep{Maercker2016}. To better understand these effects in the population of carbon stars, the mass-loss histories of additional objects need to be studied. Observatories such as SOFIA, the Atacama Large Millimeter Array (ALMA), and the James Webb Space Telescope (JWST) are well suited for these types of studies and will have a significant impact on our understanding of carbon stars and TP behavior in these objects.

\vspace{3mm}
\emph{Acknowledgments} We would like to thank the rest of the FORCAST team, George Gull, Justin Schoenwald, Chuck Henderson, Joe Adams, the USRA Science and Mission Ops teams, and the entire SOFIA staff. We would also like to thank the anonymous referee for the useful comments and suggestions on this paper. This work is based on observations made with the NASA/DLR Stratospheric Observatory for Infrared Astronomy (SOFIA). SOFIA science mission operations are conducted jointly by the Universities Space Research Association, Inc. (USRA), under NASA contract NAS2-97001, and the Deutsches SOFIA Institut (DSI) under DLR contract 50 OK 0901. Financial support for FORCAST was provided by NASA through award 8500-98-014 issued by USRA. This material is based upon work supported by the National Science Foundation Graduate Research Fellowship under Grant No. DGE-1144153. MM acknowledges financial support from the Swedish Research Council under grant number 2016-03402.

\bibliography{main}

\newpage

\begin{table*}[htp]

\vskip 3cm

\centering 
\caption{\ Observation Details}
\label{tab:tab1}
\begin{tabular}{cccccc}

\hline
Flight \# & Altitude (ft.) & Object & Filter 1 & Filter 2 & Integration (s)\\
\hline
\hline

323 & $\sim$43,000 & R Scl & 19.7 & 34.8 & 620 \\ 
323 & $\sim$43,000 & R Scl & 25.2 & 31.5 & 790 \\  

324 & $\sim$43,000 & R Scl & 25.2 & 31.5 & 270 \\
324 & $\sim$43,000 & R Scl & open & 37.1 & 1690 \\

\hline
\hline
\end{tabular}
\\ 
\centering
\end{table*}

\begin{table*}[htp]
\centering
\caption{\ DUSTY Model Parameters}
\label{tab:tab2}
\begin{tabular}{cccc}

\hline
Parameter & Value & Type & Source \\
\hline
\hline

Stellar Temperature & 2600 K & adopted & \cite{Cruzalebes2013}\\ 
Stellar Luminosity & $10^{4.0\pm0.2}~L_{\odot}$ & fitted & this work \\
Distance &  370 pc & adopted & \cite{Maercker2016} \\

Dust Size Distribution & $dn/da\propto a^{-3.5}$ & adopted  & this work \\
Fractional Dust Composition & 0.86 AmC & adopted & this work \\
   & 0.10 SiC & adopted & \cite{Sacuto2011} \\
   & 0.04 MgS &  adopted & \cite{Hony2004} \\
\hline

Shell Dust Temperature ($T_{shell}$) & 75 K & adopted & \cite{Schoier2005} \\
Shell Radius  & 6950 AU  (18.8") & fitted &  this work \\
Shell Thickness & 0.1 & adopted &  \cite{GonzalezDelgado2003} \\
Shell Dust Mass & $3.7 \times 10^{-5} M_{\odot}$ & adopted* & \cite{Schoier2005} \\
\hline

Interior Dust Temperature ($T_{in}$) & 1200 K & adopted & \cite{Sacuto2011}\\
Inner Radius ($R_{in}$) & 10 AU  (27 mas) & adopted & Determined from $T_{in}$ \\
Outer Radius ($R_{out}$) & 5550 AU  (15") & fitted & this work \\
Relative Thickness (Y) & 550 & fitted & Y=$R_{out}$/$R_{in}$ \\
Density Power Law ($\alpha$) & $0.75^{+0.45}_{-0.25}$ & fitted & this work\\

Interior Dust Mass & $9.0^{+2.3}_{-4.1}\times10^{-6}~M_{\odot}$ & fitted & this work \\

\hline
\hline
\end{tabular}

\begin{tabular}{l}
\hspace{-22mm} *For consistency, the dust shell mass has been scaled from the original value quoted in \\  \hspace{-20mm} \cite{Schoier2005} to our adopted distance of 370 pc.

\end{tabular}
\\

\centering
\end{table*}

\end{document}